\documentclass[10pt,aps,prl,twocolumn,showpacs,superscriptaddress,longbibliography]{revtex4-1}

\usepackage{graphicx}
\usepackage{amsmath}
\usepackage{amssymb}
\usepackage{bm}
\usepackage{xcolor}
\usepackage[%
  colorlinks=true,
  urlcolor=blue,
  linkcolor=blue,
  citecolor=blue
]{hyperref}
\usepackage{textgreek}

\begin{document}

\title{Non-linear Relaxation of Interacting Bosons Coherently Driven on a Narrow Optical Transition}
\author{M. Bosch Aguilera}
\affiliation{Laboratoire Kastler Brossel, Coll\`ege de France, ENS-PSL Research
University, Sorbonne Universit\'e, CNRS, 11 place Marcelin-Berthelot, 75005 Paris.}

\author{R. Bouganne}
\affiliation{Laboratoire Kastler Brossel, Coll\`ege de France, ENS-PSL Research
University, Sorbonne Universit\'e, CNRS, 11 place Marcelin-Berthelot, 75005 Paris.}

\author{A. Dareau}
\altaffiliation{Current adress: Vienna Center for Quantum Science and Technology
TU Wien – Atominstitut, Stadionallee 2, 1020 Vienna, Austria.}
\affiliation{Laboratoire Kastler Brossel, Coll\`ege de France, ENS-PSL Research
University, Sorbonne Universit\'e, CNRS, 11 place Marcelin-Berthelot, 75005 Paris.}

\author{M. Scholl}
\altaffiliation{Current adress: FARO Scanner Production GmbH, Lingwiesenstrasse 11/2, D-70825 Korntal-M{\"u}nchingen.}
\affiliation{Laboratoire Kastler Brossel, Coll\`ege de France, ENS-PSL Research
University, Sorbonne Universit\'e, CNRS, 11 place Marcelin-Berthelot, 75005 Paris.}

\author{Q. Beaufils}
\altaffiliation{Current adress: Laboratoire PhLAM, Bât. P5 - USTL, F-59655 Villeneuve d'Ascq.}
\affiliation{Laboratoire Kastler Brossel, Coll\`ege de France, ENS-PSL Research
University, Sorbonne Universit\'e, CNRS, 11 place Marcelin-Berthelot, 75005 Paris.}

\author{J. Beugnon}
\affiliation{Laboratoire Kastler Brossel, Coll\`ege de France, ENS-PSL Research
University, Sorbonne Universit\'e, CNRS, 11 place Marcelin-Berthelot, 75005 Paris.}

\author{F. Gerbier}
\email{Corresponding author: fabrice.gerbier@lkb.ens.fr}
\affiliation{Laboratoire Kastler Brossel, Coll\`ege de France, ENS-PSL Research
University, Sorbonne Universit\'e, CNRS, 11 place Marcelin-Berthelot, 75005 Paris.}


\date{\today}
\begin{abstract}
We study the dynamics of a two-component Bose-Einstein condensate (BEC) of $^{174}$Yb atoms coherently driven on a narrow optical transition. The excitation transfers the BEC to a superposition of states with different internal and momentum quantum numbers. We observe a crossover with decreasing driving strength between a regime of damped oscillations, where coherent driving prevails, and an incoherent regime, where relaxation takes over. Several relaxation mechanisms are involved: inelastic losses involving two excited atoms, leading to a non-exponential decay of populations; Doppler broadening due to the finite momentum width of the BEC and inhomogeneous elastic interactions, both leading to dephasing and to damping of the oscillations. We compare our observations to a two-component Gross-Pitaevskii (GP) model that fully includes these effects. For small or moderate densities, the damping of the oscillations is mostly due to Doppler broadening. In this regime, we find excellent agreement between the model and the experimental results. For higher densities, the role of interactions increases and so does the damping rate of the oscillations. The damping in the GP model is less pronounced than in the experiment, possibly a hint for many-body effects not captured by the mean-field description.
\end{abstract}

\maketitle

In the recent years, ultranarrow optical ``clock'' transitions interrogated by lasers with sub-Hertz frequency stability have enabled dramatic progress in time-frequency metrology \cite{ludlow2015a}. The very small radiative linewidth (low spontaneous emission rate) characterizing such transitions opens many unprecedented opportunities, {e.g.} for quantum information processing \cite{stock2008a,gorshkov2010a,daley2011a,shibata2014a}, to reach new regimes in quantum optics \cite{bohnet2012a,norcia2016a}, or to simulate complex many-body systems such as high-spin magnetism or impurity problems \cite{gorshkov2009a,martin2013a,riegger2018a}. Moreover, the recoil effect -- the increase of the atomic momentum upon absorbing a laser photon -- couples the motional state of the atoms to their internal state. This feature distinguishes single-photon transitions in the optical domain from hyperfine transitions in the radio-frequency or microwave domain, where the recoil is negligible. This enables in principle a fully coherent manipulation of the internal and external atomic state, with applications in atom interferometry \cite{hu2017a}, or in the realization of {artificial gauge potentials} \cite{gerbier2010a,livi2016a,kolkowitz2017a}. 

\begin{figure*}[ht!!!!!]
\begin{center}
\includegraphics[width=\textwidth]{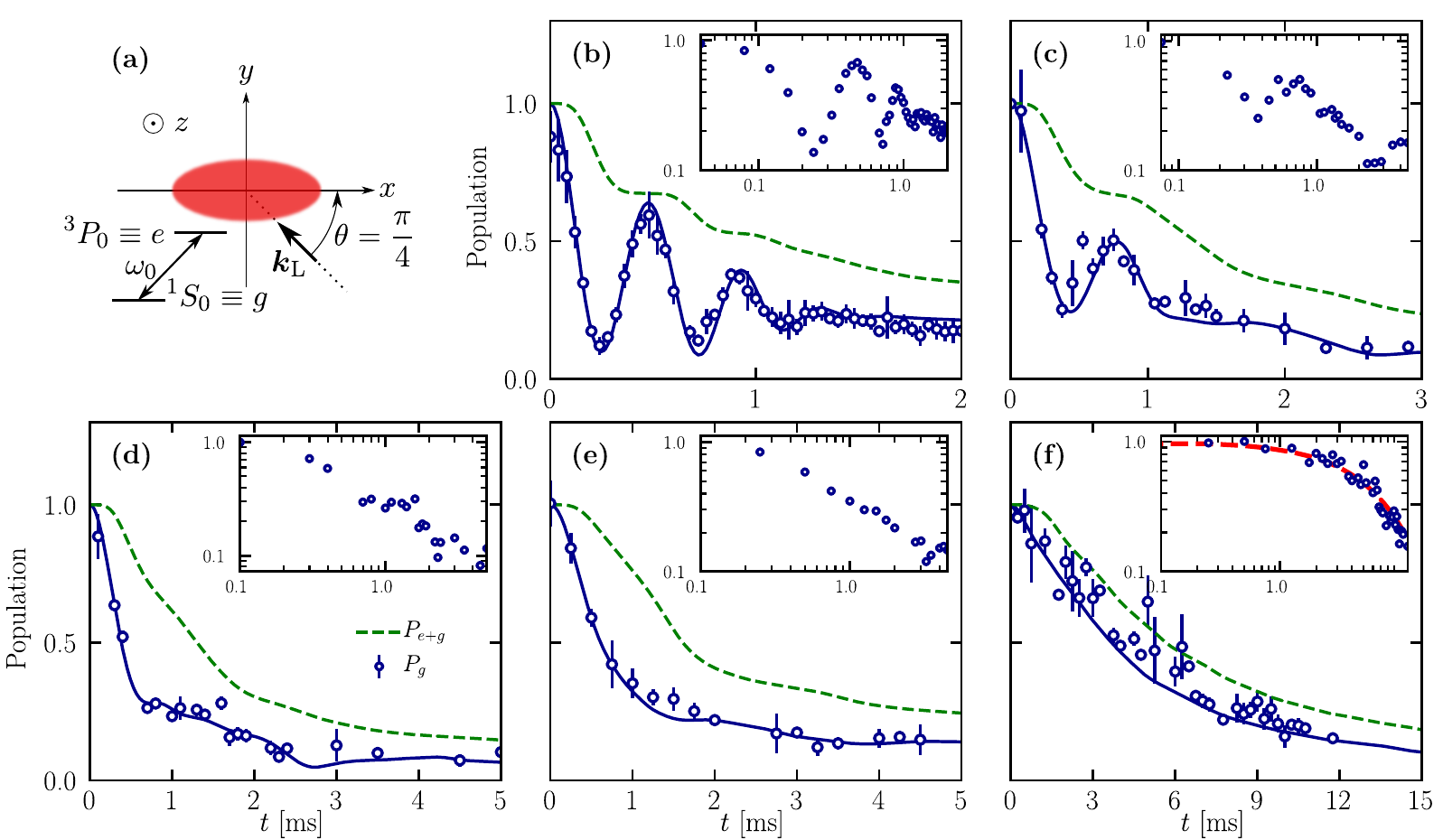}
\caption{\textbf{(a):} Sketch of the experiment\,: $^{174}$Yb atoms are probed on the clock transition connecting the ground state $g \equiv ^{1}$$S_0$ and the metastable excited state $e \equiv ^{3}$$P_0$. \textbf{(b)-(f):} Population dynamics as a function of pulse duration $t$ for varying Rabi frequency: $\Omega_{\mathrm{L}}/(2\pi) \simeq 2.1\,$kHz\,(\textbf{b}), $1.10\,$kHz\,(\textbf{c}), $750\,$Hz\,(\textbf{d}), $540\,$Hz\,(\textbf{e}) and $200\,$Hz\,(\textbf{f}). 
The circles show the measured population in $g$ normalized to the initial atom number, noted $P_g$. The solid blue lines show fits to the lossy GP model developed in the text, with only the driving strength $\Omega_{\mathrm{L}}$, initial atom number and detuning $\delta_{\mathrm{L}}'$ as free parameters. The green dashed lines show the evolution of the total atom number normalized to the initial one, noted $P_{e+g}$, according to the same model. The insets show the same data in double-logarithmic scale. The red dashed line in \textbf{(f)} shows an exponential fit to the data with a $1/\mathrm{e}$ decay rate of $\simeq{150}\,$s${}^{-1}$. For all data shown in this figure, the trap frequencies are $(\omega_x,\omega_y,\omega_z)\simeq 2\pi\times(20,264,275)\,$Hz and the BEC chemical potential is $\mu/h \simeq 1\,$kHz. } 
\label{fig.1}
\end{center}
\end{figure*}

In many of these applications, interatomic interactions play an essential role. In atomic clocks, interactions limit the clock accuracy and their role has been studied extensively \cite{ludlow2015a}. Even for fermions, where one would \textit{a priori} expect vanishing clock shifts at low temperatures, interactions lead to tiny clock shifts because of inhomogeneous excitation \cite{campbell2009a}. While atomic clocks usually operate far from quantum degeneracy, new phenomena appear in quantum degenerate gases due to the interplay between quantum statistics, the quantized motion of atoms and intra- and inter-state interactions. Optical spectroscopy has been instrumental to reveal Bose-Einstein condensation of spin-polarized atomic hydrogen \cite{fried1998a,killian1998a,killian2000a}. These experiments were performed in a weak coupling, irreversible regime suitable for spectroscopy. Still, the experimental results have not been fully understood \cite{fried1998a,bradley2001a,oktel2002a,landhuis2003a}. More recently,  one-photon spectroscopy on ultra-narrow optical transitions with spontaneous linewidth $\ll 1\,$ Hz has been reported and used to probe interaction shifts in an Yb Bose-Einstein condensate (BEC) \cite{yamaguchi2010a,notermans2016} or a Sr degenerate Fermi gas \cite{campbell2017a,marti2018a}, to measure scattering properties of fermionic\,\cite{scazza2014a,capellini2014a,zhang2015a,pagano2015a,hoefer2015a} or bosonic Yb atoms\,\cite{bouganne2017a,franchi2017a}, to study the superfluid-Mott insulator transition in an optical lattice \cite{kato2012a}, or to reveal the change in the density of states in spin-orbit coupled Fermi gases \cite{livi2016a,kolkowitz2017a}.

%
%

In this article, we report on a study of the dynamics of a BEC of $^{174}$Yb atoms coherently driven on such a narrow transition. The excitation coherently transfers the BEC in a superposition of states with different internal and momentum quantum numbers. The coherent excitation competes against a number of relaxation processes, including linear dephasing due to the finite initial momentum width and non-linear interactions, in particular inelastic processes involving two excited atoms. We observe a crossover with decreasing driving strength from a regime of damped oscillations, where coherent driving prevails, to an incoherent regime, where relaxation takes over. Throughout the crossover (except for very small driving strength), the populations relax in time with a non-exponential law. We compare our observations to a two-component Gross-Pitaevskii (GP) model that fully includes elastic and inelastic interactions and atomic motion. We find excellent agreement between the GP model and the experiment for densities around or below $10^{14}\,$at/cm$^3$, but also that the model underestimates the damping of the coherent oscillations for higher densities. This could point to additional effects beyond the GP description at play in the experiments.

We produce nearly pure BECs of $^{174}$Yb atoms in an optical crossed dipole trap (CDT) \cite{dareau2015a}. The CDT operates at the so-called magic wavelength $\lambda_{\mathrm{m}} \simeq 759.4\,$nm, where the light shifts of the electronic $^{1}S_0$ ground state and of the metastable $^{3}P_0$ excited state (denoted respectively by $g$ and $e$ in the following) are almost equal. The trapping potential of the CDT is then almost independent of the internal state. A laser near-resonant with the $g$-$e$ transition couples the two internal states (see \cite{dareau2015a} for more details on the optical setup and on the frequency stabilization). After preparing a BEC in the $g$ state, we illuminate the sample with a pulse of duration $t$, with a coupling strength $\Omega_{\mathrm{L}}$ and detuning $\delta_{\mathrm{L}} = \omega_{\mathrm{L}}-\omega_{eg}$ from the bare atomic resonance frequency $\omega_{eg}$, with $\omega_{\mathrm{L}}$ the laser frequency. The coupling laser propagates in the horizontal $x-y$ plane, with a wavevector $\bm{k}_{\mathrm{L}}$ making an angle $\theta=\pi/4$ with the weak axis ($x$-axis) of the trap (see Fig.\,\ref{fig.1}\textbf{a}). We switch off the CDT immediately after the pulse, let the cloud expand for a time of flight of $12\,$ms and record an absorption image of the atoms in $g$ using the dipole-allowed $^{1}S_0 \to$ $^{1}P_1$ transition. In the following, we focus on the normalized population $P_g$ in $g$, that is the atom number deduced from absorption images normalized to the initial one.


Fig.\,\ref{fig.1}\textbf{b}-\textbf{f} show the time evolution of $P_g$ after the coupling laser is turned on, for various Rabi frequencies. We observe damped, Rabi-like oscillations with a contrast that decreases when increasing the Rabi frequency. For all data shown in Fig.\,\ref{fig.1}, the initial condensate contains typically $\sim 10^4$ atoms for a chemical potential $\mu/h \simeq  1\,$kHz. The laser detuning is fixed to the value where we observe maximum transfer after a given pulse time (see Supplemental Material for more details). The observation of oscillations shows that the condensate is coherently transfered in a quantum superposition of $g$ and $e$. For a uniform condensate, the transfer would couple two single quantum states $\vert g, \bm{0} \rangle$ and $\vert e, \bm{k}_{\mathrm{L}} \rangle$ with momenta $ \bm{0}$ and $\hbar\bm{k}_{\mathrm{L}}$, respectively. For our sample of finite size, the two states correspond to two wavepackets centered around the same momenta and with a width $\sim\hbar/R$, where $R$ is a typical condensate size. 

Our observations are reminiscent of the behavior of two-state quantum systems, both coherently driven and incoherently coupled to a bath, such as the paradigmatic two-level atom of quantum optics or a driven qubit undergoing relaxation. {In these examples, under the assumption of short memory of the bath, one expects for weak driving an exponential decay analogous to the Wigner-Weisskopf (W-W) desintegration of a discrete level into a continuum of states\,\cite{cct_photons_atoms}.} 
The transition from W-W desintegration to underdamped oscillations with increasing driving strength can be estimated by comparing the spectral width $\Delta$ of the bath (the inverse of its memory time) to the coupling strength $\Omega_{\mathrm{L}}$: Underdamped oscillations take place in the strong driving regime $\Omega_{\mathrm{L}} \gg \Delta$ and W-W desintegration in the weak coupling regime $\Omega_{\mathrm{L}} \ll \Delta$, with a continuous change from one regime to the other. 
The same conclusions hold for an ensemble of independent two-level systems, where inhomogeneities in the coupling strength or detuning also lead to dephasing between the different members of the ensemble. This induces an additional decay of the $g$-$e$ coherence when considering ensemble-averaged quantities, translating to a reduced contrast of the oscillations. 

In the absence of interactions or of atomic motion, the damping of the oscillations would purely be due to ensemble dephasing. One form of dephasing comes from spatial inhomogeneity of the coupling strength or of the detuning. We estimate in \cite{bouganne2017a} a dephasing time of several tens of milliseconds for our experimental paramters, much longer than observed here. Another dephasing mechanism arising from ensemble averaging is Doppler broadening due to the small, but finite momentum width of the BEC. The detuning $\delta_{\mathrm{L}}' -\bm{v}_{\mathrm{R}} \cdot \bm{k}_{\mathrm{L}}$ depends on the atomic momentum $\hbar\bm{k}_{\mathrm{L}}$ due to the Doppler effect. Here $\bm{v}_{\mathrm{R}}=\hbar \bm{k}_{\mathrm{L}}/M$ is the recoil velocity, $M$ is the atomic mass, $\delta_{\textrm{L}}'=\delta_{\textrm{L}}-E_{\mathrm{R}}$ is the detuning from the recoil-shifted resonance, and $E_{\mathrm{R}}=M \bm{v}_{\mathrm{R}}^2/2$ is the recoil energy. For our experimental geometry, the finite momentum width $\sim\hbar/R_y$, with $R_y$ the size of the condensate in the most confined direction, then translates into a Doppler broadening of the resonance by $\Delta_{\mathrm{D}} = v_{\mathrm{R}}/(\sqrt{2}R)\sim 2\pi \times 600$\,Hz (see Supplemental Material for a more detailed discussion). The Doppler width $\Delta_{\mathrm{D}}$ plays the role of the spectral width, and oscillations in Fig.\,\ref{fig.1}\textbf{b}-\textbf{f} are indeed observed when $\Omega_{\mathrm{L}} \geq \Delta_{\mathrm{D}}$. 

If we describe the atoms by an internal density matrix $\hat{\rho}$, with external degrees of freedom integrated out, Doppler broadening leads to a decay of the off-diagonal elements $\rho_{eg}=\langle e \vert  \hat{\rho} \vert g \rangle$ on a time $\tau_2^\ast \sim\Delta_{\mathrm{D}}^{-1}$. For weak driving strength, assuming the damping can be accounted for by $\mathrm{d} \rho_{eg} /\mathrm{d}t = -\rho_{eg}/\tau_2^\ast $ and performing adiabatic elimination of the off-diagonal elements\,\cite{cct_photons_atoms}, one finds that the slowly-evolving population $P_g=\langle g \vert  \hat{\rho} \vert g \rangle$ decays exponentially at a rate $\Omega_{\mathrm{L}}^2\tau_2^\ast/2 \propto \Omega_{\mathrm{L}}^2/\Delta_{\mathrm{D}}$. This exponential behavior is observed for the weakest coupling used in our experiment (Fig.\,\ref{fig.1}\textbf{f}), but not for larger driving strengths where we find instead a much slower algebraic decay at long times (insets of Fig.\,\ref{fig.1}\textbf{b}-\textbf{e}). Moreover, the normalized populations in Fig.\,\ref{fig.1} do not settle to the value $1/2$ that would be expected from ensemble averaging of different momentum classes. Hence, the simple picture of the driven two-component BEC as a collection of Doppler-broadened, independent two-level systems is not sufficient to fully explain our experimental observations. 

The algebraic  decay can be ascribed to inelastic two-body losses due to principal quantum number changing collisions between two excited atoms (the rate for inelastic processes involving one ground and one excited atom is negligible \cite{bouganne2017a,franchi2017a}). Due to inelastic losses, the spatial density $\rho_e(\bm{r})$ in state $e$ decays according to

\begin{align}\label{eq.1}
\dot{\rho}_e\vert_{\textrm{inel}} = - \beta_{ee} \rho_e^2,
\end{align}
with $\beta_{ee}$ a two-body inelastic rate constant. In cases where $\rho_e \propto N$ (for instance, a uniform system prepared in $e$ and in the absence of driving), the total atom number $N$ obeys a similar equation and decays according to

\begin{align}\label{eq.2}
\frac{N(t)}{N(0)} = \frac{1}{1+{t}/{\tau_1}}
\end{align}
with a relaxation time $\tau_1 \propto 1/(\beta \rho_e)$. We find that this decay law is compatible with our observations (insets of Fig.\,\ref{fig.1}\textbf{b}-\textbf{e}).

\begin{figure}[ht]
\centering
\includegraphics[]{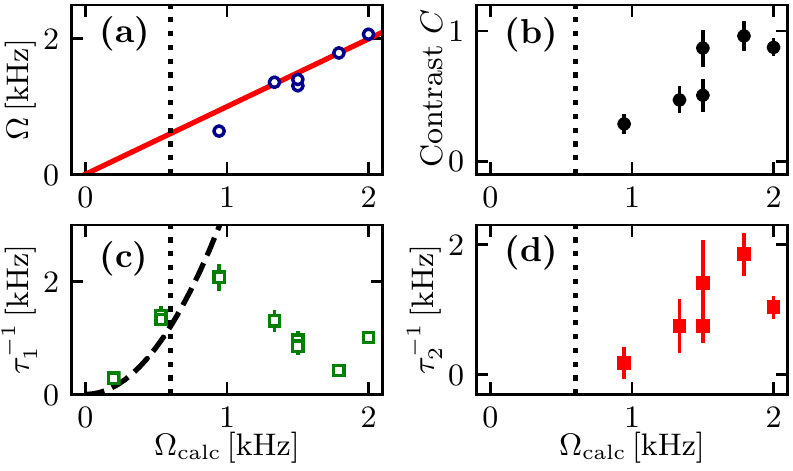}
\caption{
\textbf{(a):} Measured oscillation frequency $\Omega$, \textbf{(b):} contrast of the oscillations $C$, \textbf{(c):} amplitude damping rate $\tau_1^{-1}$, and \textbf{(d):} coherence damping rate $\tau_2^{-1}$, versus expected Rabi frequency $\Omega_{\textrm{calc}}$. The solid line in \textbf{(a)} shows the expected oscillation frequency for vanishing detuning. The dashed line in \textbf{(c)} shows $2\Omega_{\textrm{calc}}^2/\Delta_{\mathrm{D}}$. The expected decay rate for weak driving strength varies as $\Omega_{\textrm{calc}}^2/\Delta_{\mathrm{D}}$. The factor of 2 was chosen to match the first two data points. We have enforced $C=0$ in the fitting procedure for $\Omega_{\textrm{calc}} \lesssim 2\pi \times 600\,$Hz (dotted lines), where no oscillation is visible.}
\label{fig.2}
\end{figure}

To describe the crossover more quantitatively we have fitted an empirical function of the form

\begin{align}
\label{eq.3}
P_g (t) = A(t) \left(1+C \cos(\Omega t)\, \mathrm{e}^{-\frac{t}{\tau_2}} \right)
\end{align}
to the data. We chose $A(t)\propto(1+t/\tau_1)^{-1}$ for the amplitude damping function following the preceding discussion, and an exponential damping of coherences for simplicity\footnote{Other choices intead of the exponential function return a similar behavior for the fit parameters versus Rabi frequency.}. The parameters $\Omega$, $C$ and $\tau_2$ are the angular frequency, contrast and damping time of the oscillations. The best fit parameters are shown in Fig.\,\ref{fig.2} versus the expected\footnote{We compute $\Omega_{\textrm{calc}}$ from the formula given in \cite{taichenachev2006a, barber2006a}. The applied magnetic field enabling the coupling on the otherwise ``doubly forbidden" transition is $B \simeq 180\,$G. The laser waist $w \simeq 40\,$\textmu m is calculated from Gaussian beam propagation and the laser power is measured for each experiment.} Rabi frequency $\Omega_{\textrm{calc}}$. We find that the measured oscillation frequencies agree well with the expected ones (Fig.\,\ref{fig.2}\textbf{a}). Fig.\,\ref{fig.2}\textbf{b} shows how the contrast $C$ of the oscillations decreases with decreasing Rabi frequency, terminating below $\Omega_{\textrm{calc}} \lesssim 2\pi \times 600\,$Hz. 
 The inverse population and coherence damping times are also shown in Fig.\,\ref{fig.2}\textbf{c} and \textbf{d}, respectively. The threshold in Fig.\,\ref{fig.2}\textbf{b} coincides with $\Omega_{\textrm{calc}} \simeq \Delta_{\mathrm{D}}$, as expected from the picture of an ensemble of Doppler-broadened two-level systems previously discussed. The same picture explains the trend observed for weak coupling, where the effective amplitude damping rate scales as $\tau_1^{-1}\propto \Omega^2/\Delta_{\mathrm{D}}$ (dashed line in Fig.\,\ref{fig.2}\textbf{c}).

To go beyond this empirical analysis, we analyze the experimental data with a set of two GP equations describing two coherently-coupled interacting Bose gases with non-Hermitian evolution,

\begin{align}
\label{eq.4}
\mathrm{i}\hbar \frac{\partial \psi_g}{\partial t} =&\left[\hat{h} + g_{gg} \rho_g +g_{ge}  \rho_e  \right]\psi_{g}+\frac{\hbar\Omega_{\mathrm{L}}}{2}\tilde{\psi}_{e},\\
\nonumber
\mathrm{i}\hbar \frac{\partial \tilde{\psi}_{e}}{\partial t} =&\left[ \hat{h} +\bm{v}_{\mathrm{R}}\cdot \bm{\hat{p}} -\hbar\delta_{\mathrm{L}}' +g_{ee}\rho_e +g_{ge} \rho_g\right] \tilde{\psi}_{e}\\
\label{eq.5}
&-\frac{\mathrm{i}\hbar\beta_{ee}}{2}\rho_e \tilde{\psi}_{e} +\frac{\hbar\Omega_{\mathrm{L}}}{2}\psi_{g}.
\end{align}
Interactions between two ultracold atoms occupying states $\alpha$ and $\beta$ are modeled by contact potentials \cite{pitaevskii2003a} with coupling constants $g_{\alpha \beta}$ related to the $s-$wave scattering length $a_{\alpha \beta}$ by $g_{\alpha \beta}=4\pi\hbar^2 a_{\alpha \beta}/M$. For $^{174}$Yb, $a_{gg} \simeq 5.55\,$nm is accurately known from photoassociation spectroscopy \cite{kitagawa2008a}, and other elastic and inelastic scattering parameters have been measured recently using isolated atom pairs or triples in deep optical lattices \cite{bouganne2017a,franchi2017a}. Inelastic losses are taken into account by the imaginary term $\propto\beta_{ee}$. In this work, we use the most accurate measurements, namely $a_{ge} \simeq 0.9 \, a_{gg}$, $a_{ee} \simeq 1.2 \, a_{gg}$ and $\beta_{ee}\simeq 2.6\times 10^{-11}\,$cm$^3$/s \cite{bouganne2017a,franchi2017a}. The spatial densities in $g/e$ are given by $\rho_{g/e}(\bm{r}) =\vert \psi_{g/e}\vert^2$, and we have defined the single-particle Hamiltonian $\hat{h}=\bm{\hat{p}}^2/(2M)+ V_{\rm tr}$, with $\bm{\hat{p}}=-\mathrm{i} \hbar \bm{\nabla}$ the momentum operator, $V_{\rm tr}$ the harmonic trapping potential, $\tilde{\psi}_{e} = \psi_{e}\exp({-\mathrm{i} \bm{k}_{\mathrm{L}} \cdot \bm{r}})  $. The lossy GP equations (\ref{eq.4},\ref{eq.5}) derive from a master equation treated in the mean-field approximation (see Supplementary Material), and take into account all effects discussed so far -- coherent driving, intra- and inter-state interactions, coupling between internal state dynamics, atomic motion by the Doppler term $\bm{v}_{\mathrm{R}}\cdot \bm{\hat{p}}$ and inelastic losses. Interactions, losses and internal-motional coupling are of the same order of magnitude (a few hundred Hz) for our experimental parameters. 

\begin{figure}[h]
\centering
\includegraphics[]{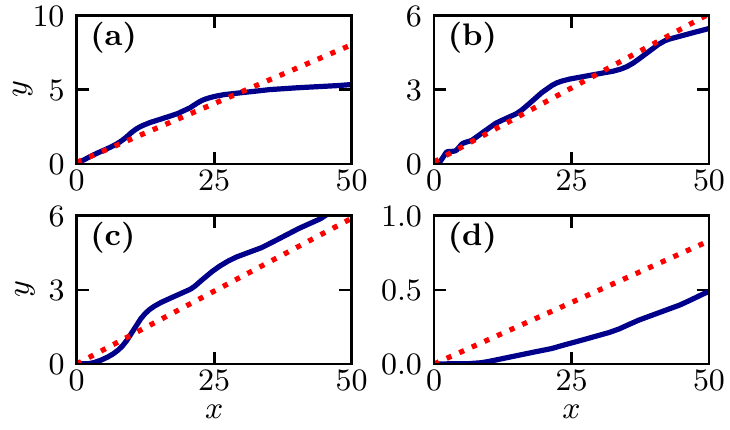}
\caption{Evolution of the normalized atom number $P_{e+g}$ versus time calculated with the lossy Gross-Pitaevskii model (solid blue line). The panels correspond to $\Omega_{\mathrm{L}}/(2\pi)=$ \textbf{(a):} $20\,$kHz, \textbf{(b):} $2\,$kHz, \textbf{(c):} $0.5\,$kHz, and \textbf{(d):} $0.1\,$kHz. By plotting $y=1/P_{e+g}-1$ versus $x=\beta_{ee}\rho_0 t$, we verify that the atom number decays as $(1+t/\tau_1)^{-1}$ for short times, with $\tau_1$ a fit parameter (dotted red line). The calculation was performed for $\delta_{\mathrm{L}}'=0$ and the parameters of the experiment shown in Fig.\,\ref{fig.1}, corresponding to a chemical potential $\mu/h=1\,$kHz. }  \label{fig.3}
\end{figure}

We solve eqs.\,(\ref{eq.4},\ref{eq.5}) numerically (see Supplemental Material) and fit the numerical solution to the experimental data with the initial atom number, coupling strength $\Omega_{\mathrm{L}}$ and detuning $\delta_{\mathrm{L}}'$ as free parameters. For all data shown in Fig.\,\ref{fig.1}, we find a good agreement between the predicted evolution of the coherently-coupled lossy GP model with the observed dynamics. The fitted Rabi frequencies are close to the expected ones (less than $10\,$\% difference), and the fitted detunings are compatible with our accuracy in finding the resonance (see Supplementary Material). 

To obtain more insight on the dynamics described by the dissipative GP equations, we simplify the experimental situation and consider a uniform system of linear size $R$ and density $\rho$. Neglecting elastic interactions and the Doppler term, we are interested in the competition between the coherent driving and the inelastic losses in the limit $\Omega_{\mathrm{L}} \gg \beta_{ee}\rho$. We then expect Rabi oscillations to develop, with the spatial densities in $e$ and $g$ given by $\rho_{e}(t)\simeq \rho(t) \sin^2(\Omega_{\mathrm{L}}t/2)$ and $\rho_{g}(t)\simeq \rho(t) \cos^2(\Omega_{\mathrm{L}}t/2)$. The envelope $\rho(t)$ slowly decays because of the inelastic losses according to eq.\,(\ref{eq.1}). After averaging over one Rabi cycle and integrating the resulting equation, we find that the cycle-averaged population $\bar{P}_g$ obeys eq.\,(\ref{eq.2}) with $\tau_1^{-1}=3\beta_{ee}\rho/4$. The expected dynamics for $\Omega_{\mathrm{L}} \geq \Delta_{\mathrm{D}}$ is thus underdamped Rabi oscillations around an average value decaying algebraically, {as observed experimentally for strong driving}.

Both the experiments and the GP calculations show that the algebraic decay persists well beyond the regime of validity of the analytic model. 
This is demonstrated in Fig.\,\ref{fig.3}, where we plot $y=1/P_{g+e}-1$ versus $x=\beta_{ee}\rho_{0}t$, with $P_{g+e}$ the total atom number (normalized to the initial one) and $\rho_{0}$ the initial peak density calculated from the GP model. The quantity $y$ depends linearly on $x$ for the algebraic decay law in eq.\,(\ref{eq.2}), and grows exponentially with $x$ for an exponential decay. We find that the decay remains algebraic unless the driving becomes very small, $\Omega_{\mathrm{L}} \ll \Delta_{\mathrm{D}}$. In this last regime [Fig.\,\ref{fig.3}\textbf{d}], we recover the Doppler-broadened  model introduced earlier with exponential damping [Fig.\,\ref{fig.1}\textbf{f}]. Even for large driving strengths, the algebraic law only holds approximately and for short times. At long times, elastic interactions, strong depletion from inelastic losses and the motion in the trap can no longer be neglected. It is then not surprising that the simple law in eq.\,(\ref{eq.2}) fails to reproduce the long-times dynamics captured by the GP equations.

\begin{figure}[h]
\centering
\includegraphics[]{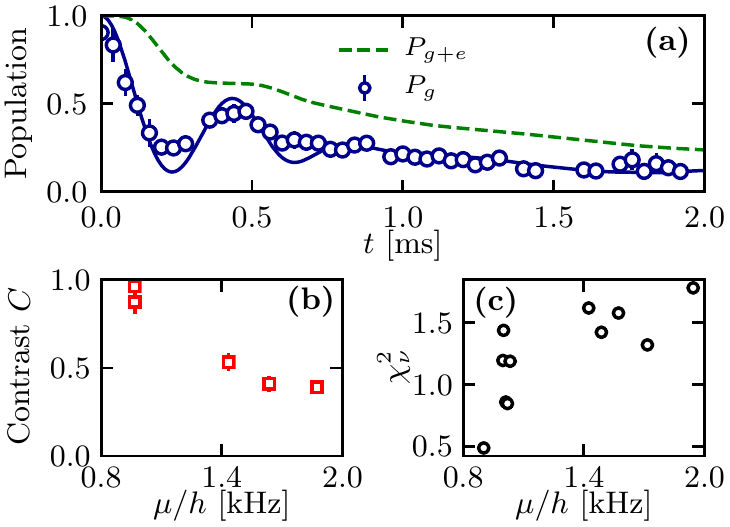}
\caption{\textbf{(a):} Population dynamics for $\mu/h \simeq 2\,$kHz. The oscillations amplitude relaxes faster than expected from the dissipative GP model, in contrast to the situation with weaker interactions ($\mu/h \simeq 1\,$kHz -- see Fig.\,\ref{fig.1}). The trap frequencies are $(\omega_x,\omega_y,\omega_z)\simeq 2\pi\times(23,570,580)\,$Hz for this measurement. \textbf{(b):} Contrast of the oscillations $C$ versus initial chemical potential $\mu$. All curves correspond to oscillations with $\Omega_{\mathrm{L}}/(2\pi) \simeq 2\,$kHz, and $\delta_{\mathrm{L}}' \simeq 0$. \textbf{(c):} Reduced $\chi^2$ of a fit to the two-component lossy GP model versus $\mu$.}
\label{fig.4}
\end{figure}

In the experiments discussed so far, relaxation of coherence or populations are mostly determined by Doppler broadening or inelastic losses, respectively, and elastic collisions are present but not essential to explain the experiments. We present in Fig.\,\ref{fig.4}\textbf{a} another set of experiments for stronger interactions ($\mu/h \sim 2\,$kHz), where elastic collisions contribute substantially to the relaxation dynamics. We find that the contrast of the oscillations, determined by the empirical fit in eq.\,(\ref{eq.3}) as before, is reduced as interactions become stronger (Fig.\,\ref{fig.4}\textbf{b}). The fit to the two-component GP model still reproduces well the long-time decay of the population, but underestimates the damping of coherences that we observe experimentally. In Fig.\,\ref{fig.4}\textbf{c}, we quantify the agreement between the GP model and the observations by a reduced $\chi^2=(1/M) \sum_{i=1}^M [f(t_i)-P_{g}(t_i)]^2/\sigma_{g,i}^2$, i.e.,the sum of the fit residuals $f(t_i)-P_{g}(t_i)$ weighted by the standard deviation $\sigma_{g,i}$ and normalized to the number $M$ of data points. We find that the reduced $\chi^2$ increases systematically with the initial chemical potential (see Fig.\,\ref{fig.4}\textbf{c}). This indicates that effects beyond the GP description become increasingly important. One such effect is momentum relaxation by collisions of the type $\vert g,\bm{0}\rangle +\vert e,\bm{k}_{\mathrm{L}} \rangle \to \vert g,\bm{q}\rangle +\vert e,\bm{k}_{\mathrm{L}}-\bm{q}\rangle$, where the notation indicates the internal and momentum states of the two atoms before and after the collision. For a uniform gas of density $\rho$, the rate of such processes is $\gamma_{\textrm{coll}}=\rho \sigma_{ge} v_{\mathrm{R}}$ with $\sigma_{ge} =4\pi a_{ge}^2$ the collisional cross-section. We find $\gamma_{\textrm{coll}} \approx 600\,$s$^{-1}$ for the typical density $\rho \simeq 5 \times 10^{14}\,$at/cm$^3$ for the experiments in Fig.\,\ref{fig.4}\textbf{a}. This simple estimate does not account for any correlations between the particles. Additional effects, {e.g.} due to thermal population of quasi-particles in the initial state or to additional fluctuations of the fields $\psi_g$ and ${{\psi}}_e$ due to the stochastic nature of the losses, could also contribute to the relaxation of coherence.

In conclusion, we have studied the coherent dynamics of a two-component, laser-driven BEC. Whereas spontaneous emission is negligible, a number of other dephasing and relaxation processes take place. We identify three effects leading to relaxation: Doppler broadening due to the finite momentum width of the trapped BEC, inelastic losses between excited atoms, and elastic interactions. We compare our observations to a two-component GP model that includes all these effects in a mean-field approach. We find excellent agreement between the model and the experiments for moderate values of the interactions, but also that the oscillations are damped more strongly in the experiment than predicted by the model for larger interactions. The discrepancy for large interactions could point to additional effects beyond the GP description, for instance the role of quasiparticles present in the initial state due to quantum or thermal fluctuations. In the context of hydrogen spectroscopy experiments \cite{fried1998a}, it has been pointed out that taking quasiparticles into account was probably necessary to explain certain features in the spectra and to resolve apparent paradoxes in the interpretation of the data\,\cite{bradley2001a}. Although the theory is more involved for strong driving than in the weak-driving, spectroscopic regime, theoretical tools, {e.g.} classical field methods\,\cite{blakie2008a,polkovnikov2010a}, are in principle available. Comparing such a calculation with our experimental results could provide an experimental test for such time-dependent classical field simulations in three dimensions. Finally, we note that the coupled two-component BEC studied in this work can be viewed as a realization of a bulk, spin-orbit coupled BEC as accomplished in several works with bosonic alkali atoms\,\cite{goldman2014a}. The mechanisms identified in this paper will be detrimental for the stability of the spin-orbit-coupled BEC. However, they could be substantially reduced in a box-like trap with a size of a few tens of microns\,\cite{gaunt2013a}. The uniform mean-field interactions should only lead to a global energy shift, the larger size of the BEC should reduce the Doppler width, and the reduced density should result in slower inelastic losses. 

\acknowledgments
We acknowledge stimulating discussions with M. H\"ofer, S. F\"olling, R. Le Targat, J. Lodewyck, Y. Le Coq and C. Kollath. This work was supported by the ERC (Starting Grant 258521--MANYBO). LKB is a member of the DIM SIRTEQ of R\'egion Ile-de-France.


\bibliography{RabiYbBEC}

\begin{thebibliography}{45}%
\makeatletter
\providecommand \@ifxundefined [1]{%
 \@ifx{#1\undefined}
}%
\providecommand \@ifnum [1]{%
 \ifnum #1\expandafter \@firstoftwo
 \else \expandafter \@secondoftwo
 \fi
}%
\providecommand \@ifx [1]{%
 \ifx #1\expandafter \@firstoftwo
 \else \expandafter \@secondoftwo
 \fi
}%
\providecommand \natexlab [1]{#1}%
\providecommand \enquote  [1]{``#1''}%
\providecommand \bibnamefont  [1]{#1}%
\providecommand \bibfnamefont [1]{#1}%
\providecommand \citenamefont [1]{#1}%
\providecommand \href@noop [0]{\@secondoftwo}%
\providecommand \href [0]{\begingroup \@sanitize@url \@href}%
\providecommand \@href[1]{\@@startlink{#1}\@@href}%
\providecommand \@@href[1]{\endgroup#1\@@endlink}%
\providecommand \@sanitize@url [0]{\catcode `\\12\catcode `\$12\catcode
  `\&12\catcode `\#12\catcode `\^12\catcode `\_12\catcode `\%12\relax}%
\providecommand \@@startlink[1]{}%
\providecommand \@@endlink[0]{}%
\providecommand \url  [0]{\begingroup\@sanitize@url \@url }%
\providecommand \@url [1]{\endgroup\@href {#1}{\urlprefix }}%
\providecommand \urlprefix  [0]{URL }%
\providecommand \Eprint [0]{\href }%
\providecommand \doibase [0]{http://dx.doi.org/}%
\providecommand \selectlanguage [0]{\@gobble}%
\providecommand \bibinfo  [0]{\@secondoftwo}%
\providecommand \bibfield  [0]{\@secondoftwo}%
\providecommand \translation [1]{[#1]}%
\providecommand \BibitemOpen [0]{}%
\providecommand \bibitemStop [0]{}%
\providecommand \bibitemNoStop [0]{.\EOS\space}%
\providecommand \EOS [0]{\spacefactor3000\relax}%
\providecommand \BibitemShut  [1]{\csname bibitem#1\endcsname}%
\let\auto@bib@innerbib\@empty
\bibitem [{\citenamefont {Ludlow}\ \emph {et~al.}(2015)\citenamefont {Ludlow},
  \citenamefont {Boyd}, \citenamefont {Ye}, \citenamefont {Peik},\ and\
  \citenamefont {Schmidt}}]{ludlow2015a}%
  \BibitemOpen
  \bibfield  {author} {\bibinfo {author} {\bibfnamefont {A.~D.}\ \bibnamefont
  {Ludlow}}, \bibinfo {author} {\bibfnamefont {M.~M.}\ \bibnamefont {Boyd}},
  \bibinfo {author} {\bibfnamefont {J.}~\bibnamefont {Ye}}, \bibinfo {author}
  {\bibfnamefont {E.}~\bibnamefont {Peik}}, \ and\ \bibinfo {author}
  {\bibfnamefont {P.~O.}\ \bibnamefont {Schmidt}},\ }\href {\doibase
  10.1103/RevModPhys.87.637} {\bibfield  {journal} {\bibinfo  {journal} {Rev.
  Mod. Phys.}\ }\textbf {\bibinfo {volume} {87}},\ \bibinfo {pages} {637}
  (\bibinfo {year} {2015})}\BibitemShut {NoStop}%
\bibitem [{\citenamefont {Stock}\ \emph {et~al.}(2008)\citenamefont {Stock},
  \citenamefont {Babcock}, \citenamefont {Raizen},\ and\ \citenamefont
  {Sanders}}]{stock2008a}%
  \BibitemOpen
  \bibfield  {author} {\bibinfo {author} {\bibfnamefont {R.}~\bibnamefont
  {Stock}}, \bibinfo {author} {\bibfnamefont {N.~S.}\ \bibnamefont {Babcock}},
  \bibinfo {author} {\bibfnamefont {M.~G.}\ \bibnamefont {Raizen}}, \ and\
  \bibinfo {author} {\bibfnamefont {B.~C.}\ \bibnamefont {Sanders}},\
  }\href@noop {} {\bibfield  {journal} {\bibinfo  {journal} {Phys. Rev. A}\
  }\textbf {\bibinfo {volume} {78}},\ \bibinfo {eid} {022301} (\bibinfo {year}
  {2008})}\BibitemShut {NoStop}%
\bibitem [{\citenamefont {Gorshkov}\ \emph {et~al.}(2010)\citenamefont
  {Gorshkov}, \citenamefont {Hermele}, \citenamefont {Gurarie}, \citenamefont
  {Xu}, \citenamefont {Julienne}, \citenamefont {Ye}, \citenamefont {Zoller},
  \citenamefont {Demler}, \citenamefont {Lukin},\ and\ \citenamefont
  {Rey}}]{gorshkov2010a}%
  \BibitemOpen
  \bibfield  {author} {\bibinfo {author} {\bibfnamefont {A.~V.}\ \bibnamefont
  {Gorshkov}}, \bibinfo {author} {\bibfnamefont {M.}~\bibnamefont {Hermele}},
  \bibinfo {author} {\bibfnamefont {V.}~\bibnamefont {Gurarie}}, \bibinfo
  {author} {\bibfnamefont {C.}~\bibnamefont {Xu}}, \bibinfo {author}
  {\bibfnamefont {P.~S.}\ \bibnamefont {Julienne}}, \bibinfo {author}
  {\bibfnamefont {J.}~\bibnamefont {Ye}}, \bibinfo {author} {\bibfnamefont
  {P.}~\bibnamefont {Zoller}}, \bibinfo {author} {\bibfnamefont
  {E.}~\bibnamefont {Demler}}, \bibinfo {author} {\bibfnamefont {M.~D.}\
  \bibnamefont {Lukin}}, \ and\ \bibinfo {author} {\bibfnamefont {A.~M.}\
  \bibnamefont {Rey}},\ }\href@noop {} {\bibfield  {journal} {\bibinfo
  {journal} {Nat Phys}\ }\textbf {\bibinfo {volume} {6}},\ \bibinfo {pages}
  {289} (\bibinfo {year} {2010})}\BibitemShut {NoStop}%
\bibitem [{\citenamefont {Daley}(2011)}]{daley2011a}%
  \BibitemOpen
  \bibfield  {author} {\bibinfo {author} {\bibfnamefont {A.}~\bibnamefont
  {Daley}},\ }\href {\doibase 10.1007/s11128-011-0293-3} {\bibfield  {journal}
  {\bibinfo  {journal} {Quantum Information Processing}\ }\textbf {\bibinfo
  {volume} {10}},\ \bibinfo {pages} {865} (\bibinfo {year} {2011})}\BibitemShut
  {NoStop}%
\bibitem [{\citenamefont {Shibata}\ \emph {et~al.}(2014)\citenamefont
  {Shibata}, \citenamefont {Yamamoto}, \citenamefont {Seki},\ and\
  \citenamefont {Takahashi}}]{shibata2014a}%
  \BibitemOpen
  \bibfield  {author} {\bibinfo {author} {\bibfnamefont {K.}~\bibnamefont
  {Shibata}}, \bibinfo {author} {\bibfnamefont {R.}~\bibnamefont {Yamamoto}},
  \bibinfo {author} {\bibfnamefont {Y.}~\bibnamefont {Seki}}, \ and\ \bibinfo
  {author} {\bibfnamefont {Y.}~\bibnamefont {Takahashi}},\ }\href {\doibase
  10.1103/PhysRevA.89.031601} {\bibfield  {journal} {\bibinfo  {journal} {Phys.
  Rev. A}\ }\textbf {\bibinfo {volume} {89}},\ \bibinfo {pages} {031601}
  (\bibinfo {year} {2014})}\BibitemShut {NoStop}%
\bibitem [{\citenamefont {Bohnet}\ \emph {et~al.}(2012)\citenamefont {Bohnet},
  \citenamefont {Chen}, \citenamefont {Weiner}, \citenamefont {Meiser},
  \citenamefont {Holland},\ and\ \citenamefont {Thompson}}]{bohnet2012a}%
  \BibitemOpen
  \bibfield  {author} {\bibinfo {author} {\bibfnamefont {J.~G.}\ \bibnamefont
  {Bohnet}}, \bibinfo {author} {\bibfnamefont {Z.}~\bibnamefont {Chen}},
  \bibinfo {author} {\bibfnamefont {J.~M.}\ \bibnamefont {Weiner}}, \bibinfo
  {author} {\bibfnamefont {D.}~\bibnamefont {Meiser}}, \bibinfo {author}
  {\bibfnamefont {M.~J.}\ \bibnamefont {Holland}}, \ and\ \bibinfo {author}
  {\bibfnamefont {J.~K.}\ \bibnamefont {Thompson}},\ }\href@noop {} {\bibfield
  {journal} {\bibinfo  {journal} {Nature}\ }\textbf {\bibinfo {volume} {484}},\
  \bibinfo {pages} {78} (\bibinfo {year} {2012})}\BibitemShut {NoStop}%
\bibitem [{\citenamefont {Norcia}\ \emph {et~al.}(2016)\citenamefont {Norcia},
  \citenamefont {Winchester}, \citenamefont {Cline},\ and\ \citenamefont
  {Thompson}}]{norcia2016a}%
  \BibitemOpen
  \bibfield  {author} {\bibinfo {author} {\bibfnamefont {M.~A.}\ \bibnamefont
  {Norcia}}, \bibinfo {author} {\bibfnamefont {M.~N.}\ \bibnamefont
  {Winchester}}, \bibinfo {author} {\bibfnamefont {J.~R.~K.}\ \bibnamefont
  {Cline}}, \ and\ \bibinfo {author} {\bibfnamefont {J.~K.}\ \bibnamefont
  {Thompson}},\ }\href {\doibase 10.1126/sciadv.1601231} {\bibfield  {journal}
  {\bibinfo  {journal} {Science Advances}\ }\textbf {\bibinfo {volume} {2}}
  (\bibinfo {year} {2016}),\ 10.1126/sciadv.1601231}\BibitemShut {NoStop}%
\bibitem [{\citenamefont {Gorshkov}\ \emph {et~al.}(2009)\citenamefont
  {Gorshkov}, \citenamefont {Rey}, \citenamefont {Daley}, \citenamefont {Boyd},
  \citenamefont {Ye}, \citenamefont {Zoller},\ and\ \citenamefont
  {Lukin}}]{gorshkov2009a}%
  \BibitemOpen
  \bibfield  {author} {\bibinfo {author} {\bibfnamefont {A.~V.}\ \bibnamefont
  {Gorshkov}}, \bibinfo {author} {\bibfnamefont {A.~M.}\ \bibnamefont {Rey}},
  \bibinfo {author} {\bibfnamefont {A.~J.}\ \bibnamefont {Daley}}, \bibinfo
  {author} {\bibfnamefont {M.~M.}\ \bibnamefont {Boyd}}, \bibinfo {author}
  {\bibfnamefont {J.}~\bibnamefont {Ye}}, \bibinfo {author} {\bibfnamefont
  {P.}~\bibnamefont {Zoller}}, \ and\ \bibinfo {author} {\bibfnamefont {M.~D.}\
  \bibnamefont {Lukin}},\ }\href {\doibase 10.1103/PhysRevLett.102.110503}
  {\bibfield  {journal} {\bibinfo  {journal} {Phys. Rev. Lett.}\ }\textbf
  {\bibinfo {volume} {102}},\ \bibinfo {pages} {110503} (\bibinfo {year}
  {2009})}\BibitemShut {NoStop}%
\bibitem [{\citenamefont {Martin}\ \emph {et~al.}(2013)\citenamefont {Martin},
  \citenamefont {Bishof}, \citenamefont {Swallows}, \citenamefont {Zhang},
  \citenamefont {Benko}, \citenamefont {von Stecher}, \citenamefont {Gorshkov},
  \citenamefont {Rey},\ and\ \citenamefont {Ye}}]{martin2013a}%
  \BibitemOpen
  \bibfield  {author} {\bibinfo {author} {\bibfnamefont {M.~J.}\ \bibnamefont
  {Martin}}, \bibinfo {author} {\bibfnamefont {M.}~\bibnamefont {Bishof}},
  \bibinfo {author} {\bibfnamefont {M.~D.}\ \bibnamefont {Swallows}}, \bibinfo
  {author} {\bibfnamefont {X.}~\bibnamefont {Zhang}}, \bibinfo {author}
  {\bibfnamefont {C.}~\bibnamefont {Benko}}, \bibinfo {author} {\bibfnamefont
  {J.}~\bibnamefont {von Stecher}}, \bibinfo {author} {\bibfnamefont {A.~V.}\
  \bibnamefont {Gorshkov}}, \bibinfo {author} {\bibfnamefont {A.~M.}\
  \bibnamefont {Rey}}, \ and\ \bibinfo {author} {\bibfnamefont
  {J.}~\bibnamefont {Ye}},\ }\href {\doibase 10.1126/science.1236929}
  {\bibfield  {journal} {\bibinfo  {journal} {Science}\ }\textbf {\bibinfo
  {volume} {341}},\ \bibinfo {pages} {632} (\bibinfo {year}
  {2013})}\BibitemShut {NoStop}%
\bibitem [{\citenamefont {Riegger}\ \emph {et~al.}(2018)\citenamefont
  {Riegger}, \citenamefont {Darkwah~Oppong}, \citenamefont {H\"ofer},
  \citenamefont {Fernandes}, \citenamefont {Bloch},\ and\ \citenamefont
  {F\"olling}}]{riegger2018a}%
  \BibitemOpen
  \bibfield  {author} {\bibinfo {author} {\bibfnamefont {L.}~\bibnamefont
  {Riegger}}, \bibinfo {author} {\bibfnamefont {N.}~\bibnamefont
  {Darkwah~Oppong}}, \bibinfo {author} {\bibfnamefont {M.}~\bibnamefont
  {H\"ofer}}, \bibinfo {author} {\bibfnamefont {D.~R.}\ \bibnamefont
  {Fernandes}}, \bibinfo {author} {\bibfnamefont {I.}~\bibnamefont {Bloch}}, \
  and\ \bibinfo {author} {\bibfnamefont {S.}~\bibnamefont {F\"olling}},\ }\href
  {\doibase 10.1103/PhysRevLett.120.143601} {\bibfield  {journal} {\bibinfo
  {journal} {Phys. Rev. Lett.}\ }\textbf {\bibinfo {volume} {120}},\ \bibinfo
  {pages} {143601} (\bibinfo {year} {2018})}\BibitemShut {NoStop}%
\bibitem [{\citenamefont {Hu}\ \emph {et~al.}(2017)\citenamefont {Hu},
  \citenamefont {Poli}, \citenamefont {Salvi},\ and\ \citenamefont
  {Tino}}]{hu2017a}%
  \BibitemOpen
  \bibfield  {author} {\bibinfo {author} {\bibfnamefont {L.}~\bibnamefont
  {Hu}}, \bibinfo {author} {\bibfnamefont {N.}~\bibnamefont {Poli}}, \bibinfo
  {author} {\bibfnamefont {L.}~\bibnamefont {Salvi}}, \ and\ \bibinfo {author}
  {\bibfnamefont {G.~M.}\ \bibnamefont {Tino}},\ }\href {\doibase
  10.1103/PhysRevLett.119.263601} {\bibfield  {journal} {\bibinfo  {journal}
  {Phys. Rev. Lett.}\ }\textbf {\bibinfo {volume} {119}},\ \bibinfo {pages}
  {263601} (\bibinfo {year} {2017})}\BibitemShut {NoStop}%
\bibitem [{\citenamefont {Gerbier}\ and\ \citenamefont
  {Dalibard}(2010)}]{gerbier2010a}%
  \BibitemOpen
  \bibfield  {author} {\bibinfo {author} {\bibfnamefont {F.}~\bibnamefont
  {Gerbier}}\ and\ \bibinfo {author} {\bibfnamefont {J.}~\bibnamefont
  {Dalibard}},\ }\href@noop {} {\bibfield  {journal} {\bibinfo  {journal} {New
  Journal of Physics}\ }\textbf {\bibinfo {volume} {12}},\ \bibinfo {pages}
  {033007} (\bibinfo {year} {2010})}\BibitemShut {NoStop}%
\bibitem [{\citenamefont {Livi}\ \emph {et~al.}(2016)\citenamefont {Livi},
  \citenamefont {Cappellini}, \citenamefont {Diem}, \citenamefont {Franchi},
  \citenamefont {Clivati}, \citenamefont {Frittelli}, \citenamefont {Levi},
  \citenamefont {Calonico}, \citenamefont {Catani}, \citenamefont {Inguscio},\
  and\ \citenamefont {Fallani}}]{livi2016a}%
  \BibitemOpen
  \bibfield  {author} {\bibinfo {author} {\bibfnamefont {L.~F.}\ \bibnamefont
  {Livi}}, \bibinfo {author} {\bibfnamefont {G.}~\bibnamefont {Cappellini}},
  \bibinfo {author} {\bibfnamefont {M.}~\bibnamefont {Diem}}, \bibinfo {author}
  {\bibfnamefont {L.}~\bibnamefont {Franchi}}, \bibinfo {author} {\bibfnamefont
  {C.}~\bibnamefont {Clivati}}, \bibinfo {author} {\bibfnamefont
  {M.}~\bibnamefont {Frittelli}}, \bibinfo {author} {\bibfnamefont
  {F.}~\bibnamefont {Levi}}, \bibinfo {author} {\bibfnamefont {D.}~\bibnamefont
  {Calonico}}, \bibinfo {author} {\bibfnamefont {J.}~\bibnamefont {Catani}},
  \bibinfo {author} {\bibfnamefont {M.}~\bibnamefont {Inguscio}}, \ and\
  \bibinfo {author} {\bibfnamefont {L.}~\bibnamefont {Fallani}},\ }\href
  {\doibase 10.1103/PhysRevLett.117.220401} {\bibfield  {journal} {\bibinfo
  {journal} {Phys. Rev. Lett.}\ }\textbf {\bibinfo {volume} {117}},\ \bibinfo
  {pages} {220401} (\bibinfo {year} {2016})}\BibitemShut {NoStop}%
\bibitem [{\citenamefont {Kolkowitz}\ \emph {et~al.}(2017)\citenamefont
  {Kolkowitz}, \citenamefont {Bromley}, \citenamefont {Bothwell}, \citenamefont
  {Wall}, \citenamefont {Marti}, \citenamefont {Koller}, \citenamefont {Zhang},
  \citenamefont {Rey},\ and\ \citenamefont {Ye}}]{kolkowitz2017a}%
  \BibitemOpen
  \bibfield  {author} {\bibinfo {author} {\bibfnamefont {S.}~\bibnamefont
  {Kolkowitz}}, \bibinfo {author} {\bibfnamefont {S.~L.}\ \bibnamefont
  {Bromley}}, \bibinfo {author} {\bibfnamefont {T.}~\bibnamefont {Bothwell}},
  \bibinfo {author} {\bibfnamefont {M.~L.}\ \bibnamefont {Wall}}, \bibinfo
  {author} {\bibfnamefont {G.~E.}\ \bibnamefont {Marti}}, \bibinfo {author}
  {\bibfnamefont {A.~P.}\ \bibnamefont {Koller}}, \bibinfo {author}
  {\bibfnamefont {X.}~\bibnamefont {Zhang}}, \bibinfo {author} {\bibfnamefont
  {A.~M.}\ \bibnamefont {Rey}}, \ and\ \bibinfo {author} {\bibfnamefont
  {J.}~\bibnamefont {Ye}},\ }\href@noop {} {\bibfield  {journal} {\bibinfo
  {journal} {Nature}\ }\textbf {\bibinfo {volume} {542}},\ \bibinfo {pages}
  {66} (\bibinfo {year} {2017})}\BibitemShut {NoStop}%
\bibitem [{\citenamefont {Campbell}\ \emph {et~al.}(2009)\citenamefont
  {Campbell}, \citenamefont {Boyd}, \citenamefont {Thomsen}, \citenamefont
  {Martin}, \citenamefont {Blatt}, \citenamefont {Swallows}, \citenamefont
  {Nicholson}, \citenamefont {Fortier}, \citenamefont {Oates}, \citenamefont
  {Diddams}, \citenamefont {Lemke}, \citenamefont {Naidon}, \citenamefont
  {Julienne}, \citenamefont {Ye},\ and\ \citenamefont
  {Ludlow}}]{campbell2009a}%
  \BibitemOpen
  \bibfield  {author} {\bibinfo {author} {\bibfnamefont {G.~K.}\ \bibnamefont
  {Campbell}}, \bibinfo {author} {\bibfnamefont {M.~M.}\ \bibnamefont {Boyd}},
  \bibinfo {author} {\bibfnamefont {J.~W.}\ \bibnamefont {Thomsen}}, \bibinfo
  {author} {\bibfnamefont {M.~J.}\ \bibnamefont {Martin}}, \bibinfo {author}
  {\bibfnamefont {S.}~\bibnamefont {Blatt}}, \bibinfo {author} {\bibfnamefont
  {M.~D.}\ \bibnamefont {Swallows}}, \bibinfo {author} {\bibfnamefont {T.~L.}\
  \bibnamefont {Nicholson}}, \bibinfo {author} {\bibfnamefont {T.}~\bibnamefont
  {Fortier}}, \bibinfo {author} {\bibfnamefont {C.~W.}\ \bibnamefont {Oates}},
  \bibinfo {author} {\bibfnamefont {S.~A.}\ \bibnamefont {Diddams}}, \bibinfo
  {author} {\bibfnamefont {N.~D.}\ \bibnamefont {Lemke}}, \bibinfo {author}
  {\bibfnamefont {P.}~\bibnamefont {Naidon}}, \bibinfo {author} {\bibfnamefont
  {P.}~\bibnamefont {Julienne}}, \bibinfo {author} {\bibfnamefont
  {J.}~\bibnamefont {Ye}}, \ and\ \bibinfo {author} {\bibfnamefont {A.~D.}\
  \bibnamefont {Ludlow}},\ }\href {\doibase 10.1126/science.1169724} {\bibfield
   {journal} {\bibinfo  {journal} {Science}\ }\textbf {\bibinfo {volume}
  {324}},\ \bibinfo {pages} {360} (\bibinfo {year} {2009})}\BibitemShut
  {NoStop}%
\bibitem [{\citenamefont {Fried}\ \emph {et~al.}(1998)\citenamefont {Fried},
  \citenamefont {Killian}, \citenamefont {Willmann}, \citenamefont {Landhuis},
  \citenamefont {Moss}, \citenamefont {Kleppner},\ and\ \citenamefont
  {Greytak}}]{fried1998a}%
  \BibitemOpen
  \bibfield  {author} {\bibinfo {author} {\bibfnamefont {D.~G.}\ \bibnamefont
  {Fried}}, \bibinfo {author} {\bibfnamefont {T.~C.}\ \bibnamefont {Killian}},
  \bibinfo {author} {\bibfnamefont {L.}~\bibnamefont {Willmann}}, \bibinfo
  {author} {\bibfnamefont {D.}~\bibnamefont {Landhuis}}, \bibinfo {author}
  {\bibfnamefont {S.~C.}\ \bibnamefont {Moss}}, \bibinfo {author}
  {\bibfnamefont {D.}~\bibnamefont {Kleppner}}, \ and\ \bibinfo {author}
  {\bibfnamefont {T.~J.}\ \bibnamefont {Greytak}},\ }\href {\doibase
  10.1103/PhysRevLett.81.3811} {\bibfield  {journal} {\bibinfo  {journal}
  {Phys. Rev. Lett.}\ }\textbf {\bibinfo {volume} {81}},\ \bibinfo {pages}
  {3811} (\bibinfo {year} {1998})}\BibitemShut {NoStop}%
\bibitem [{\citenamefont {Killian}\ \emph {et~al.}(1998)\citenamefont
  {Killian}, \citenamefont {Fried}, \citenamefont {Willmann}, \citenamefont
  {Landhuis}, \citenamefont {Moss}, \citenamefont {Greytak},\ and\
  \citenamefont {Kleppner}}]{killian1998a}%
  \BibitemOpen
  \bibfield  {author} {\bibinfo {author} {\bibfnamefont {T.~C.}\ \bibnamefont
  {Killian}}, \bibinfo {author} {\bibfnamefont {D.~G.}\ \bibnamefont {Fried}},
  \bibinfo {author} {\bibfnamefont {L.}~\bibnamefont {Willmann}}, \bibinfo
  {author} {\bibfnamefont {D.}~\bibnamefont {Landhuis}}, \bibinfo {author}
  {\bibfnamefont {S.~C.}\ \bibnamefont {Moss}}, \bibinfo {author}
  {\bibfnamefont {T.~J.}\ \bibnamefont {Greytak}}, \ and\ \bibinfo {author}
  {\bibfnamefont {D.}~\bibnamefont {Kleppner}},\ }\href {\doibase
  10.1103/PhysRevLett.81.3807} {\bibfield  {journal} {\bibinfo  {journal}
  {Phys. Rev. Lett.}\ }\textbf {\bibinfo {volume} {81}},\ \bibinfo {pages}
  {3807} (\bibinfo {year} {1998})}\BibitemShut {NoStop}%
\bibitem [{\citenamefont {Killian}(2000)}]{killian2000a}%
  \BibitemOpen
  \bibfield  {author} {\bibinfo {author} {\bibfnamefont {T.~C.}\ \bibnamefont
  {Killian}},\ }\href {\doibase 10.1103/PhysRevA.61.033611} {\bibfield
  {journal} {\bibinfo  {journal} {Phys. Rev. A}\ }\textbf {\bibinfo {volume}
  {61}},\ \bibinfo {pages} {033611} (\bibinfo {year} {2000})}\BibitemShut
  {NoStop}%
\bibitem [{\citenamefont {Gardiner}\ and\ \citenamefont
  {Bradley}(2001)}]{bradley2001a}%
  \BibitemOpen
  \bibfield  {author} {\bibinfo {author} {\bibfnamefont {C.~W.}\ \bibnamefont
  {Gardiner}}\ and\ \bibinfo {author} {\bibfnamefont {A.~S.}\ \bibnamefont
  {Bradley}},\ }\href@noop {} {\bibfield  {journal} {\bibinfo  {journal}
  {Journal of Physics B: Atomic, Molecular and Optical Physics}\ }\textbf
  {\bibinfo {volume} {34}},\ \bibinfo {pages} {4663} (\bibinfo {year}
  {2001})}\BibitemShut {NoStop}%
\bibitem [{\citenamefont {Oktel}\ \emph {et~al.}(2002)\citenamefont {Oktel},
  \citenamefont {Killian}, \citenamefont {Kleppner},\ and\ \citenamefont
  {Levitov}}]{oktel2002a}%
  \BibitemOpen
  \bibfield  {author} {\bibinfo {author} {\bibfnamefont {M.~O.}\ \bibnamefont
  {Oktel}}, \bibinfo {author} {\bibfnamefont {T.~C.}\ \bibnamefont {Killian}},
  \bibinfo {author} {\bibfnamefont {D.}~\bibnamefont {Kleppner}}, \ and\
  \bibinfo {author} {\bibfnamefont {L.~S.}\ \bibnamefont {Levitov}},\ }\href
  {\doibase 10.1103/PhysRevA.65.033617} {\bibfield  {journal} {\bibinfo
  {journal} {Phys. Rev. A}\ }\textbf {\bibinfo {volume} {65}},\ \bibinfo
  {pages} {033617} (\bibinfo {year} {2002})}\BibitemShut {NoStop}%
\bibitem [{\citenamefont {Landhuis}\ \emph {et~al.}(2003)\citenamefont
  {Landhuis}, \citenamefont {Matos}, \citenamefont {Moss}, \citenamefont
  {Steinberger}, \citenamefont {Vant}, \citenamefont {Willmann}, \citenamefont
  {Greytak},\ and\ \citenamefont {Kleppner}}]{landhuis2003a}%
  \BibitemOpen
  \bibfield  {author} {\bibinfo {author} {\bibfnamefont {D.}~\bibnamefont
  {Landhuis}}, \bibinfo {author} {\bibfnamefont {L.}~\bibnamefont {Matos}},
  \bibinfo {author} {\bibfnamefont {S.~C.}\ \bibnamefont {Moss}}, \bibinfo
  {author} {\bibfnamefont {J.~K.}\ \bibnamefont {Steinberger}}, \bibinfo
  {author} {\bibfnamefont {K.}~\bibnamefont {Vant}}, \bibinfo {author}
  {\bibfnamefont {L.}~\bibnamefont {Willmann}}, \bibinfo {author}
  {\bibfnamefont {T.~J.}\ \bibnamefont {Greytak}}, \ and\ \bibinfo {author}
  {\bibfnamefont {D.}~\bibnamefont {Kleppner}},\ }\href {\doibase
  10.1103/PhysRevA.67.022718} {\bibfield  {journal} {\bibinfo  {journal} {Phys.
  Rev. A}\ }\textbf {\bibinfo {volume} {67}},\ \bibinfo {pages} {022718}
  (\bibinfo {year} {2003})}\BibitemShut {NoStop}%
\bibitem [{\citenamefont {Yamaguchi}\ \emph {et~al.}(2010)\citenamefont
  {Yamaguchi}, \citenamefont {Uetake}, \citenamefont {Kato}, \citenamefont
  {Ito},\ and\ \citenamefont {Takahashi}}]{yamaguchi2010a}%
  \BibitemOpen
  \bibfield  {author} {\bibinfo {author} {\bibfnamefont {A.}~\bibnamefont
  {Yamaguchi}}, \bibinfo {author} {\bibfnamefont {S.}~\bibnamefont {Uetake}},
  \bibinfo {author} {\bibfnamefont {S.}~\bibnamefont {Kato}}, \bibinfo {author}
  {\bibfnamefont {H.}~\bibnamefont {Ito}}, \ and\ \bibinfo {author}
  {\bibfnamefont {Y.}~\bibnamefont {Takahashi}},\ }\href@noop {} {\bibfield
  {journal} {\bibinfo  {journal} {New Journal of Physics}\ }\textbf {\bibinfo
  {volume} {12}},\ \bibinfo {pages} {103001} (\bibinfo {year}
  {2010})}\BibitemShut {NoStop}%
\bibitem [{\citenamefont {Notermans}\ \emph {et~al.}(2016)\citenamefont
  {Notermans}, \citenamefont {Rengelink},\ and\ \citenamefont
  {Vassen}}]{notermans2016}%
  \BibitemOpen
  \bibfield  {author} {\bibinfo {author} {\bibfnamefont {R.~P. M. J.~W.}\
  \bibnamefont {Notermans}}, \bibinfo {author} {\bibfnamefont {R.~J.}\
  \bibnamefont {Rengelink}}, \ and\ \bibinfo {author} {\bibfnamefont
  {W.}~\bibnamefont {Vassen}},\ }\href {\doibase
  10.1103/PhysRevLett.117.213001} {\bibfield  {journal} {\bibinfo  {journal}
  {Phys. Rev. Lett.}\ }\textbf {\bibinfo {volume} {117}},\ \bibinfo {pages}
  {213001} (\bibinfo {year} {2016})}\BibitemShut {NoStop}%
\bibitem [{\citenamefont {Campbell}\ \emph {et~al.}(2017)\citenamefont
  {Campbell}, \citenamefont {Hutson}, \citenamefont {Marti}, \citenamefont
  {Goban}, \citenamefont {Darkwah~Oppong}, \citenamefont {McNally},
  \citenamefont {Sonderhouse}, \citenamefont {Robinson}, \citenamefont {Zhang},
  \citenamefont {Bloom},\ and\ \citenamefont {Ye}}]{campbell2017a}%
  \BibitemOpen
  \bibfield  {author} {\bibinfo {author} {\bibfnamefont {S.~L.}\ \bibnamefont
  {Campbell}}, \bibinfo {author} {\bibfnamefont {R.~B.}\ \bibnamefont
  {Hutson}}, \bibinfo {author} {\bibfnamefont {G.~E.}\ \bibnamefont {Marti}},
  \bibinfo {author} {\bibfnamefont {A.}~\bibnamefont {Goban}}, \bibinfo
  {author} {\bibfnamefont {N.}~\bibnamefont {Darkwah~Oppong}}, \bibinfo
  {author} {\bibfnamefont {R.~L.}\ \bibnamefont {McNally}}, \bibinfo {author}
  {\bibfnamefont {L.}~\bibnamefont {Sonderhouse}}, \bibinfo {author}
  {\bibfnamefont {J.~M.}\ \bibnamefont {Robinson}}, \bibinfo {author}
  {\bibfnamefont {W.}~\bibnamefont {Zhang}}, \bibinfo {author} {\bibfnamefont
  {B.~J.}\ \bibnamefont {Bloom}}, \ and\ \bibinfo {author} {\bibfnamefont
  {J.}~\bibnamefont {Ye}},\ }\href {\doibase 10.1126/science.aam5538}
  {\bibfield  {journal} {\bibinfo  {journal} {Science}\ }\textbf {\bibinfo
  {volume} {358}},\ \bibinfo {pages} {90} (\bibinfo {year} {2017})}\BibitemShut
  {NoStop}%
\bibitem [{\citenamefont {Marti}\ \emph {et~al.}(2018)\citenamefont {Marti},
  \citenamefont {Hutson}, \citenamefont {Goban}, \citenamefont {Campbell},
  \citenamefont {Poli},\ and\ \citenamefont {Ye}}]{marti2018a}%
  \BibitemOpen
  \bibfield  {author} {\bibinfo {author} {\bibfnamefont {G.~E.}\ \bibnamefont
  {Marti}}, \bibinfo {author} {\bibfnamefont {R.~B.}\ \bibnamefont {Hutson}},
  \bibinfo {author} {\bibfnamefont {A.}~\bibnamefont {Goban}}, \bibinfo
  {author} {\bibfnamefont {S.~L.}\ \bibnamefont {Campbell}}, \bibinfo {author}
  {\bibfnamefont {N.}~\bibnamefont {Poli}}, \ and\ \bibinfo {author}
  {\bibfnamefont {J.}~\bibnamefont {Ye}},\ }\href {\doibase
  10.1103/PhysRevLett.120.103201} {\bibfield  {journal} {\bibinfo  {journal}
  {Phys. Rev. Lett.}\ }\textbf {\bibinfo {volume} {120}},\ \bibinfo {pages}
  {103201} (\bibinfo {year} {2018})}\BibitemShut {NoStop}%
\bibitem [{\citenamefont {Scazza}\ \emph {et~al.}(2014)\citenamefont {Scazza},
  \citenamefont {Hofrichter}, \citenamefont {Hofer}, \citenamefont {De~Groot},
  \citenamefont {Bloch},\ and\ \citenamefont {Folling}}]{scazza2014a}%
  \BibitemOpen
  \bibfield  {author} {\bibinfo {author} {\bibfnamefont {F.}~\bibnamefont
  {Scazza}}, \bibinfo {author} {\bibfnamefont {C.}~\bibnamefont {Hofrichter}},
  \bibinfo {author} {\bibfnamefont {M.}~\bibnamefont {Hofer}}, \bibinfo
  {author} {\bibfnamefont {P.~C.}\ \bibnamefont {De~Groot}}, \bibinfo {author}
  {\bibfnamefont {I.}~\bibnamefont {Bloch}}, \ and\ \bibinfo {author}
  {\bibfnamefont {S.}~\bibnamefont {Folling}},\ }\href@noop {} {\bibfield
  {journal} {\bibinfo  {journal} {Nat Phys}\ }\textbf {\bibinfo {volume}
  {10}},\ \bibinfo {pages} {779} (\bibinfo {year} {2014})}\BibitemShut
  {NoStop}%
\bibitem [{\citenamefont {Cappellini}\ \emph {et~al.}(2014)\citenamefont
  {Cappellini}, \citenamefont {Mancini}, \citenamefont {Pagano}, \citenamefont
  {Lombardi}, \citenamefont {Livi}, \citenamefont {Siciliani~de Cumis},
  \citenamefont {Cancio}, \citenamefont {Pizzocaro}, \citenamefont {Calonico},
  \citenamefont {Levi}, \citenamefont {Sias}, \citenamefont {Catani},
  \citenamefont {Inguscio},\ and\ \citenamefont {Fallani}}]{capellini2014a}%
  \BibitemOpen
  \bibfield  {author} {\bibinfo {author} {\bibfnamefont {G.}~\bibnamefont
  {Cappellini}}, \bibinfo {author} {\bibfnamefont {M.}~\bibnamefont {Mancini}},
  \bibinfo {author} {\bibfnamefont {G.}~\bibnamefont {Pagano}}, \bibinfo
  {author} {\bibfnamefont {P.}~\bibnamefont {Lombardi}}, \bibinfo {author}
  {\bibfnamefont {L.}~\bibnamefont {Livi}}, \bibinfo {author} {\bibfnamefont
  {M.}~\bibnamefont {Siciliani~de Cumis}}, \bibinfo {author} {\bibfnamefont
  {P.}~\bibnamefont {Cancio}}, \bibinfo {author} {\bibfnamefont
  {M.}~\bibnamefont {Pizzocaro}}, \bibinfo {author} {\bibfnamefont
  {D.}~\bibnamefont {Calonico}}, \bibinfo {author} {\bibfnamefont
  {F.}~\bibnamefont {Levi}}, \bibinfo {author} {\bibfnamefont {C.}~\bibnamefont
  {Sias}}, \bibinfo {author} {\bibfnamefont {J.}~\bibnamefont {Catani}},
  \bibinfo {author} {\bibfnamefont {M.}~\bibnamefont {Inguscio}}, \ and\
  \bibinfo {author} {\bibfnamefont {L.}~\bibnamefont {Fallani}},\ }\href
  {\doibase 10.1103/PhysRevLett.113.120402} {\bibfield  {journal} {\bibinfo
  {journal} {Phys. Rev. Lett.}\ }\textbf {\bibinfo {volume} {113}},\ \bibinfo
  {pages} {120402} (\bibinfo {year} {2014})}\BibitemShut {NoStop}%
\bibitem [{\citenamefont {Zhang}\ \emph {et~al.}(2015)\citenamefont {Zhang},
  \citenamefont {Cheng}, \citenamefont {Zhai},\ and\ \citenamefont
  {Zhang}}]{zhang2015a}%
  \BibitemOpen
  \bibfield  {author} {\bibinfo {author} {\bibfnamefont {R.}~\bibnamefont
  {Zhang}}, \bibinfo {author} {\bibfnamefont {Y.}~\bibnamefont {Cheng}},
  \bibinfo {author} {\bibfnamefont {H.}~\bibnamefont {Zhai}}, \ and\ \bibinfo
  {author} {\bibfnamefont {P.}~\bibnamefont {Zhang}},\ }\href {\doibase
  10.1103/PhysRevLett.115.135301} {\bibfield  {journal} {\bibinfo  {journal}
  {Phys. Rev. Lett.}\ }\textbf {\bibinfo {volume} {115}},\ \bibinfo {pages}
  {135301} (\bibinfo {year} {2015})}\BibitemShut {NoStop}%
\bibitem [{\citenamefont {Pagano}\ \emph {et~al.}(2015)\citenamefont {Pagano},
  \citenamefont {Mancini}, \citenamefont {Cappellini}, \citenamefont {Livi},
  \citenamefont {Sias}, \citenamefont {Catani}, \citenamefont {Inguscio},\ and\
  \citenamefont {Fallani}}]{pagano2015a}%
  \BibitemOpen
  \bibfield  {author} {\bibinfo {author} {\bibfnamefont {G.}~\bibnamefont
  {Pagano}}, \bibinfo {author} {\bibfnamefont {M.}~\bibnamefont {Mancini}},
  \bibinfo {author} {\bibfnamefont {G.}~\bibnamefont {Cappellini}}, \bibinfo
  {author} {\bibfnamefont {L.}~\bibnamefont {Livi}}, \bibinfo {author}
  {\bibfnamefont {C.}~\bibnamefont {Sias}}, \bibinfo {author} {\bibfnamefont
  {J.}~\bibnamefont {Catani}}, \bibinfo {author} {\bibfnamefont
  {M.}~\bibnamefont {Inguscio}}, \ and\ \bibinfo {author} {\bibfnamefont
  {L.}~\bibnamefont {Fallani}},\ }\href {\doibase
  10.1103/PhysRevLett.115.265301} {\bibfield  {journal} {\bibinfo  {journal}
  {Phys. Rev. Lett.}\ }\textbf {\bibinfo {volume} {115}},\ \bibinfo {pages}
  {265301} (\bibinfo {year} {2015})}\BibitemShut {NoStop}%
\bibitem [{\citenamefont {H\"ofer}\ \emph {et~al.}(2015)\citenamefont
  {H\"ofer}, \citenamefont {Riegger}, \citenamefont {Scazza}, \citenamefont
  {Hofrichter}, \citenamefont {Fernandes}, \citenamefont {Parish},
  \citenamefont {Levinsen}, \citenamefont {Bloch},\ and\ \citenamefont
  {F\"olling}}]{hoefer2015a}%
  \BibitemOpen
  \bibfield  {author} {\bibinfo {author} {\bibfnamefont {M.}~\bibnamefont
  {H\"ofer}}, \bibinfo {author} {\bibfnamefont {L.}~\bibnamefont {Riegger}},
  \bibinfo {author} {\bibfnamefont {F.}~\bibnamefont {Scazza}}, \bibinfo
  {author} {\bibfnamefont {C.}~\bibnamefont {Hofrichter}}, \bibinfo {author}
  {\bibfnamefont {D.~R.}\ \bibnamefont {Fernandes}}, \bibinfo {author}
  {\bibfnamefont {M.~M.}\ \bibnamefont {Parish}}, \bibinfo {author}
  {\bibfnamefont {J.}~\bibnamefont {Levinsen}}, \bibinfo {author}
  {\bibfnamefont {I.}~\bibnamefont {Bloch}}, \ and\ \bibinfo {author}
  {\bibfnamefont {S.}~\bibnamefont {F\"olling}},\ }\href {\doibase
  10.1103/PhysRevLett.115.265302} {\bibfield  {journal} {\bibinfo  {journal}
  {Phys. Rev. Lett.}\ }\textbf {\bibinfo {volume} {115}},\ \bibinfo {pages}
  {265302} (\bibinfo {year} {2015})}\BibitemShut {NoStop}%
\bibitem [{\citenamefont {Bouganne}\ \emph {et~al.}(2017)\citenamefont
  {Bouganne}, \citenamefont {{Bosch Aguilera}}, \citenamefont {Dareau},
  \citenamefont {Soave}, \citenamefont {Beugnon},\ and\ \citenamefont
  {Gerbier}}]{bouganne2017a}%
  \BibitemOpen
  \bibfield  {author} {\bibinfo {author} {\bibfnamefont {R.}~\bibnamefont
  {Bouganne}}, \bibinfo {author} {\bibfnamefont {M.}~\bibnamefont {{Bosch
  Aguilera}}}, \bibinfo {author} {\bibfnamefont {A.}~\bibnamefont {Dareau}},
  \bibinfo {author} {\bibfnamefont {E.}~\bibnamefont {Soave}}, \bibinfo
  {author} {\bibfnamefont {J.}~\bibnamefont {Beugnon}}, \ and\ \bibinfo
  {author} {\bibfnamefont {F.}~\bibnamefont {Gerbier}},\ }\href@noop {}
  {\bibfield  {journal} {\bibinfo  {journal} {New Journal of Physics}\ }\textbf
  {\bibinfo {volume} {19}},\ \bibinfo {pages} {113006} (\bibinfo {year}
  {2017})}\BibitemShut {NoStop}%
\bibitem [{\citenamefont {{Franchi}}\ \emph {et~al.}(2017)\citenamefont
  {{Franchi}}, \citenamefont {{Livi}}, \citenamefont {{Cappellini}},
  \citenamefont {{Binella}}, \citenamefont {{Inguscio}}, \citenamefont
  {{Catani}},\ and\ \citenamefont {{Fallani}}}]{franchi2017a}%
  \BibitemOpen
  \bibfield  {author} {\bibinfo {author} {\bibfnamefont {L.}~\bibnamefont
  {{Franchi}}}, \bibinfo {author} {\bibfnamefont {L.~F.}\ \bibnamefont
  {{Livi}}}, \bibinfo {author} {\bibfnamefont {G.}~\bibnamefont
  {{Cappellini}}}, \bibinfo {author} {\bibfnamefont {G.}~\bibnamefont
  {{Binella}}}, \bibinfo {author} {\bibfnamefont {M.}~\bibnamefont
  {{Inguscio}}}, \bibinfo {author} {\bibfnamefont {J.}~\bibnamefont
  {{Catani}}}, \ and\ \bibinfo {author} {\bibfnamefont {L.}~\bibnamefont
  {{Fallani}}},\ }\href@noop {} {\bibfield  {journal} {\bibinfo  {journal} {New
  J. Phys.}\ }\textbf {\bibinfo {volume} {19}},\ \bibinfo {pages} {103037}
  (\bibinfo {year} {2017})}\BibitemShut {NoStop}%
\bibitem [{\citenamefont {Kato}\ \emph {et~al.}(2012)\citenamefont {Kato},
  \citenamefont {Shibata}, \citenamefont {Yamamoto}, \citenamefont
  {Yoshikawa},\ and\ \citenamefont {Takahashi}}]{kato2012a}%
  \BibitemOpen
  \bibfield  {author} {\bibinfo {author} {\bibfnamefont {S.}~\bibnamefont
  {Kato}}, \bibinfo {author} {\bibfnamefont {K.}~\bibnamefont {Shibata}},
  \bibinfo {author} {\bibfnamefont {R.}~\bibnamefont {Yamamoto}}, \bibinfo
  {author} {\bibfnamefont {Y.}~\bibnamefont {Yoshikawa}}, \ and\ \bibinfo
  {author} {\bibfnamefont {Y.}~\bibnamefont {Takahashi}},\ }\href {\doibase
  10.1007/s00340-012-4893-0} {\bibfield  {journal} {\bibinfo  {journal}
  {Applied Physics B}\ }\textbf {\bibinfo {volume} {108}},\ \bibinfo {pages}
  {31} (\bibinfo {year} {2012})}\BibitemShut {NoStop}%
\bibitem [{\citenamefont {Dareau}\ \emph {et~al.}(2015)\citenamefont {Dareau},
  \citenamefont {Scholl}, \citenamefont {Beaufils}, \citenamefont {D\"oring},
  \citenamefont {Beugnon},\ and\ \citenamefont {Gerbier}}]{dareau2015a}%
  \BibitemOpen
  \bibfield  {author} {\bibinfo {author} {\bibfnamefont {A.}~\bibnamefont
  {Dareau}}, \bibinfo {author} {\bibfnamefont {M.}~\bibnamefont {Scholl}},
  \bibinfo {author} {\bibfnamefont {Q.}~\bibnamefont {Beaufils}}, \bibinfo
  {author} {\bibfnamefont {D.}~\bibnamefont {D\"oring}}, \bibinfo {author}
  {\bibfnamefont {J.}~\bibnamefont {Beugnon}}, \ and\ \bibinfo {author}
  {\bibfnamefont {F.}~\bibnamefont {Gerbier}},\ }\href {\doibase
  10.1103/PhysRevA.91.023626} {\bibfield  {journal} {\bibinfo  {journal} {Phys.
  Rev. A}\ }\textbf {\bibinfo {volume} {91}},\ \bibinfo {pages} {023626}
  (\bibinfo {year} {2015})}\BibitemShut {NoStop}%
\bibitem [{\citenamefont {Cohen-Tannoudji}\ \emph {et~al.}(1997)\citenamefont
  {Cohen-Tannoudji}, \citenamefont {Dupont-Roc},\ and\ \citenamefont
  {Grynberg}}]{cct_photons_atoms}%
  \BibitemOpen
  \bibfield  {author} {\bibinfo {author} {\bibfnamefont {C.}~\bibnamefont
  {Cohen-Tannoudji}}, \bibinfo {author} {\bibfnamefont {J.}~\bibnamefont
  {Dupont-Roc}}, \ and\ \bibinfo {author} {\bibfnamefont {G.}~\bibnamefont
  {Grynberg}},\ }\href@noop {} {}\ (\bibinfo  {publisher} {Wiley VCH},\
  \bibinfo {address} {New York},\ \bibinfo {year} {1997})\BibitemShut {NoStop}%
\bibitem [{Note1()}]{Note1}%
  \BibitemOpen
  \bibinfo {note} {Other choices intead of the exponential function return a
  similar behavior for the fit parameters versus Rabi frequency.}\BibitemShut
  {Stop}%
\bibitem [{Note2()}]{Note2}%
  \BibitemOpen
  \bibinfo {note} {We compute $\Omega _{\protect \textrm {calc}}$ from the
  formula given in \cite {taichenachev2006a, barber2006a}. The applied magnetic
  field enabling the coupling on the otherwise ``doubly forbidden}\BibitemShut
  {NoStop}%
\bibitem [{\citenamefont {Pitaevskii}\ and\ \citenamefont
  {Stringari}(2003)}]{pitaevskii2003a}%
  \BibitemOpen
  \bibfield  {author} {\bibinfo {author} {\bibfnamefont {L.}~\bibnamefont
  {Pitaevskii}}\ and\ \bibinfo {author} {\bibfnamefont {S.}~\bibnamefont
  {Stringari}},\ }\href@noop {} {\emph {\bibinfo {title} {Bose Einstein
  condensation}}}\ (\bibinfo  {publisher} {Oxford University Press},\ \bibinfo
  {address} {Oxford},\ \bibinfo {year} {2003})\BibitemShut {NoStop}%
\bibitem [{\citenamefont {Kitagawa}\ \emph {et~al.}(2008)\citenamefont
  {Kitagawa}, \citenamefont {Enomoto}, \citenamefont {Kasa}, \citenamefont
  {Takahashi}, \citenamefont {Ciurylo}, \citenamefont {Naidon},\ and\
  \citenamefont {Julienne}}]{kitagawa2008a}%
  \BibitemOpen
  \bibfield  {author} {\bibinfo {author} {\bibfnamefont {M.}~\bibnamefont
  {Kitagawa}}, \bibinfo {author} {\bibfnamefont {K.}~\bibnamefont {Enomoto}},
  \bibinfo {author} {\bibfnamefont {K.}~\bibnamefont {Kasa}}, \bibinfo {author}
  {\bibfnamefont {Y.}~\bibnamefont {Takahashi}}, \bibinfo {author}
  {\bibfnamefont {R.}~\bibnamefont {Ciurylo}}, \bibinfo {author} {\bibfnamefont
  {P.}~\bibnamefont {Naidon}}, \ and\ \bibinfo {author} {\bibfnamefont {P.~S.}\
  \bibnamefont {Julienne}},\ }\href {\doibase 10.1103/PhysRevA.77.012719}
  {\bibfield  {journal} {\bibinfo  {journal} {Phys. Rev. A}\ }\textbf {\bibinfo
  {volume} {77}},\ \bibinfo {eid} {012719} (\bibinfo {year}
  {2008})}\BibitemShut {NoStop}%
\bibitem [{\citenamefont {Blakie}\ \emph {et~al.}(2008)\citenamefont {Blakie},
  \citenamefont {Bradley}, \citenamefont {Davis}, \citenamefont {Ballagh},\
  and\ \citenamefont {Gardiner}}]{blakie2008a}%
  \BibitemOpen
  \bibfield  {author} {\bibinfo {author} {\bibfnamefont {P.}~\bibnamefont
  {Blakie}}, \bibinfo {author} {\bibfnamefont {A.}~\bibnamefont {Bradley}},
  \bibinfo {author} {\bibfnamefont {M.}~\bibnamefont {Davis}}, \bibinfo
  {author} {\bibfnamefont {R.}~\bibnamefont {Ballagh}}, \ and\ \bibinfo
  {author} {\bibfnamefont {C.}~\bibnamefont {Gardiner}},\ }\href {\doibase
  10.1080/00018730802564254} {\bibfield  {journal} {\bibinfo  {journal}
  {Advances in Physics}\ }\textbf {\bibinfo {volume} {57}},\ \bibinfo {pages}
  {363} (\bibinfo {year} {2008})}\BibitemShut {NoStop}%
\bibitem [{\citenamefont {Polkovnikov}(2010)}]{polkovnikov2010a}%
  \BibitemOpen
  \bibfield  {author} {\bibinfo {author} {\bibfnamefont {A.}~\bibnamefont
  {Polkovnikov}},\ }\href@noop {} {\bibfield  {journal} {\bibinfo  {journal}
  {Annals of Physics}\ }\textbf {\bibinfo {volume} {325}} (\bibinfo {year}
  {2010})}\BibitemShut {NoStop}%
\bibitem [{\citenamefont {Goldman}\ \emph {et~al.}(2014)\citenamefont
  {Goldman}, \citenamefont {Juzeliūnas}, \citenamefont {Öhberg},\ and\
  \citenamefont {Spielman}}]{goldman2014a}%
  \BibitemOpen
  \bibfield  {author} {\bibinfo {author} {\bibfnamefont {N.}~\bibnamefont
  {Goldman}}, \bibinfo {author} {\bibfnamefont {G.}~\bibnamefont
  {Juzeliūnas}}, \bibinfo {author} {\bibfnamefont {P.}~\bibnamefont
  {Öhberg}}, \ and\ \bibinfo {author} {\bibfnamefont {I.~B.}\ \bibnamefont
  {Spielman}},\ }\href {http://stacks.iop.org/0034-4885/77/i=12/a=126401}
  {\bibfield  {journal} {\bibinfo  {journal} {Reports on Progress in Physics}\
  }\textbf {\bibinfo {volume} {77}},\ \bibinfo {pages} {126401} (\bibinfo
  {year} {2014})}\BibitemShut {NoStop}%
\bibitem [{\citenamefont {Gaunt}\ \emph {et~al.}(2013)\citenamefont {Gaunt},
  \citenamefont {Schmidutz}, \citenamefont {Gotlibovych}, \citenamefont
  {Smith},\ and\ \citenamefont {Hadzibabic}}]{gaunt2013a}%
  \BibitemOpen
  \bibfield  {author} {\bibinfo {author} {\bibfnamefont {A.~L.}\ \bibnamefont
  {Gaunt}}, \bibinfo {author} {\bibfnamefont {T.~F.}\ \bibnamefont
  {Schmidutz}}, \bibinfo {author} {\bibfnamefont {I.}~\bibnamefont
  {Gotlibovych}}, \bibinfo {author} {\bibfnamefont {R.~P.}\ \bibnamefont
  {Smith}}, \ and\ \bibinfo {author} {\bibfnamefont {Z.}~\bibnamefont
  {Hadzibabic}},\ }\href {\doibase 10.1103/PhysRevLett.110.200406} {\bibfield
  {journal} {\bibinfo  {journal} {Phys. Rev. Lett.}\ }\textbf {\bibinfo
  {volume} {110}},\ \bibinfo {pages} {200406} (\bibinfo {year}
  {2013})}\BibitemShut {NoStop}%
\bibitem [{\citenamefont {Taichenachev}\ \emph {et~al.}(2006)\citenamefont
  {Taichenachev}, \citenamefont {Yudin}, \citenamefont {Oates}, \citenamefont
  {Hoyt}, \citenamefont {Barber},\ and\ \citenamefont
  {Hollberg}}]{taichenachev2006a}%
  \BibitemOpen
  \bibfield  {author} {\bibinfo {author} {\bibfnamefont {A.~V.}\ \bibnamefont
  {Taichenachev}}, \bibinfo {author} {\bibfnamefont {V.~I.}\ \bibnamefont
  {Yudin}}, \bibinfo {author} {\bibfnamefont {C.~W.}\ \bibnamefont {Oates}},
  \bibinfo {author} {\bibfnamefont {C.~W.}\ \bibnamefont {Hoyt}}, \bibinfo
  {author} {\bibfnamefont {Z.~W.}\ \bibnamefont {Barber}}, \ and\ \bibinfo
  {author} {\bibfnamefont {L.}~\bibnamefont {Hollberg}},\ }\href {\doibase
  10.1103/PhysRevLett.96.083001} {\bibfield  {journal} {\bibinfo  {journal}
  {Phys. Rev. Lett.}\ }\textbf {\bibinfo {volume} {96}},\ \bibinfo {pages}
  {083001} (\bibinfo {year} {2006})}\BibitemShut {NoStop}%
\bibitem [{\citenamefont {Barber}\ \emph {et~al.}(2006)\citenamefont {Barber},
  \citenamefont {Hoyt}, \citenamefont {Oates}, \citenamefont {Hollberg},
  \citenamefont {Taichenachev},\ and\ \citenamefont {Yudin}}]{barber2006a}%
  \BibitemOpen
  \bibfield  {author} {\bibinfo {author} {\bibfnamefont {Z.~W.}\ \bibnamefont
  {Barber}}, \bibinfo {author} {\bibfnamefont {C.~W.}\ \bibnamefont {Hoyt}},
  \bibinfo {author} {\bibfnamefont {C.~W.}\ \bibnamefont {Oates}}, \bibinfo
  {author} {\bibfnamefont {L.}~\bibnamefont {Hollberg}}, \bibinfo {author}
  {\bibfnamefont {A.~V.}\ \bibnamefont {Taichenachev}}, \ and\ \bibinfo
  {author} {\bibfnamefont {V.~I.}\ \bibnamefont {Yudin}},\ }\href {\doibase
  10.1103/PhysRevLett.96.083002} {\bibfield  {journal} {\bibinfo  {journal}
  {Phys. Rev. Lett.}\ }\textbf {\bibinfo {volume} {96}},\ \bibinfo {pages}
  {083002} (\bibinfo {year} {2006})}\BibitemShut {NoStop}%
\end{thebibliography}%
\bibliographystyle{apsrev_nourl}

\end{document}